\documentclass[16pt, conference]{IEEEtran}
\IEEEoverridecommandlockouts
\usepackage{cite}
\usepackage{amsmath,amssymb,amsfonts}
\usepackage{algorithmic}
\usepackage{graphicx}
\usepackage{textcomp}
\usepackage{xcolor}
\usepackage{dblfloatfix}
\usepackage[normalem]{ulem}
\usepackage{graphicx}
\usepackage{caption}
\usepackage{multirow}
\usepackage[hyphens]{url}
\usepackage{hyperref}
\useunder{\uline}{\ul}{}
\def\BibTeX{{\rm B\kern-.05em{\sc i\kern-.025em b}\kern-.08em
    T\kern-.1667em\lower.7ex\hbox{E}\kern-.125emX}}
\makeatletter
\def\normalsize{\@setfontsize{\normalsize}{13bp}{15.00pt}}
\normalsize
\makeatother
\begin{document}

\title{AndroScanner: Automated Backend Vulnerability Detection for Android Applications\\

}

\author{\IEEEauthorblockA{\textit{Harini Dandu}}
\textit{hdandu3@gatech.edu} \\
\textit{Master of Science in Cybersecurity} \\
\textit{Georgia Institute of Technology}\\

}

\maketitle
% \begin{figure*}[h]
% \centering
%   \includegraphics[width=0.7\textwidth,keepaspectratio]{Workflow.png}
%   \caption{Workflow of BEE}
% \end{figure*}

\begin{abstract}
Nowadays, there exists a mobile application for every feature from ordering a coffee to monitoring bank accounts. Backends at these mobile applications provide various functionalities such as content delivery, analytics, ad networks, telemetry, and more for the daily maintenance of the apps. These functionalities are mainly delivered by the APIs. Unfortunately, application developers are unaware or have no control over the security practices when choosing or managing these services. The mobile applications using these backends are in turn installed by millions of users potentially affecting thousands of them. These exposed backend vulnerabilities can be used by attackers to compromise the mobile backend, which can result in leaking user data, deleting application content, or injecting malicious code. It is necessary to analyze these mobile apps' backends and provide actionable remediation to the application developers. This paper presents \textit{AndroScanner}, an automated pipeline for vetting the backends that mobile applications interact with and providing actionable remedies. We evaluate \textit{AndroScanner} on two Android applications: a purposely vulnerable bank application and \textit{Hirect}, a production recruitment application with over 50k downloads on Google Play Store. Across both applications, \textit{AndroScanner} extracted 24 APIs (4 and 20 respectively), identified 5 total vulnerabilities (4 in the bank app and 1 in a recruitment app), and discovered a previously unreported Excessive Data Exposure vulnerability in the recruitment application ranked 3rd in the OWASP API Security Top 10. The tool is available upon request to assist app developers in improving the security of their mobile backends.
\vspace{1pt}

KEYWORDS: Android Security, Mobile Application Security, API Vulnerability Detection, Static Analysis, Dynamic Analysis, Penetration Testing, OWASP API Security, APK Analysis, Excessive Data Exposure, Security Automation.
\end{abstract}

\section{Introduction}

More than 5.3 billion people use mobile applications\cite{1}. Around 90 percent of individuals rely on smartphones. Roughly about 40 apps exist on each of these phones. Mobile backends provide various features such as content delivery, ad networks, telemetry, and more. These features are supported by several layers of software and multiple vendors including cloud providers, hosting providers, and content delivery networks (CDNs) that offer managed platforms, operating systems, and physical/virtual hardware. The inherent complexity of these backends makes secure deployment and maintenance difficult. As a result, when selecting these infrastructures for creating or renting mobile backends, application developers frequently overlook the security standards.\cite{2}

Backend breaches of mobile applications which happened recently indicate how prevalent these attacks are. The Fortnite mobile game hijacking\cite{3} demonstrated how progressively downloaded material from mobile backends can allow an attacker to install new mobile apps without the user's knowledge.

Even if the developer is security conscious, because of third-party libraries, it is unclear with which backends their mobile app will interface. Third-party libraries do not reveal their backends to developers; instead, they provide an application program interface (API) via which developers may interact. Many of these risks are avoidable if developers have the necessary tools and resources to assess the security of their backends. Furthermore, identifying insecure software layers and the responsible party helps speed up the remediation, lowering the risk of exposure.

Unfortunately, existing solutions like Drozer\cite{4} demand practical recommendations for mobile app developers. The latest study on server-side vulnerability detection of mobile apps\cite{7,8,9} also has revealed that app developers' lack of security knowledge is a rising issue. However, by focusing solely on the software service layer of mobile backends, these studies merely scrape the surface.

To identify the most significant difficulties impacting mobile backends, a thorough analysis of APIs is required. Furthermore, in order to carry out such a study, the analysis must be reproducible, transparent, and simple for developers to comprehend. The research should be conducted on a representative mobile app ecosystem in order to offer a clear picture of the backend vulnerability environment. Finally, the research should provide vulnerabilities to follow in order to assist and inform them about the security of their mobile backends.

To the end, this paper presents the design and implementation of \textit{AndroScanner}, an automated analysis pipeline to study mobile backends. Using \textit{AndroScanner}, I have tested two applications,  Bank App which is a vulnerable application\cite{5}, and the \textit{Hirect} application\cite{6} which has over 50k+ downloads on Google Playstore. \textit{AndroScanner} retrieves a list of backend APIs from an input APK, using remote vetting techniques to discover software vulnerabilities and accountable parties, and offers a list of existing vulnerabilities to the app developer. Findings discuss the total number of vulnerabilities present in each of these applications considering different scenarios.

\section{Design}

The design for \textit{AndroScanner} can be broken down into three major parts. They are as follows:
\begin{enumerate}
 \item Extracting API calls using Static Analysis and Dynamic Analysis.
 \item Vetting the extracted APIs for vulnerabilities.
 \item Reporting the list of vulnerabilities to the target user.
\end{enumerate}

\begin{figure*}[h]
\centering
  \includegraphics[width=0.99\textwidth,keepaspectratio]{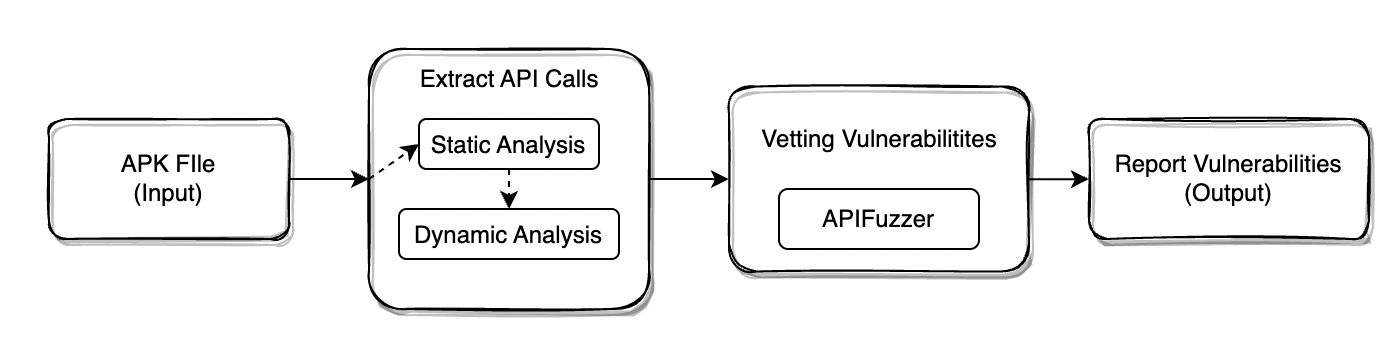}
  \caption{Workflow of AndroScanner}
  \label{design}
\end{figure*}

Figure \ref{design} depicts the whole design of the system. Let’s deep dive into each of the implementation steps.

\section{Extracting the API Calls}
The first step of the automated pipeline \textit{AndroScanner} is extracting the API calls that are involved in the backend application. As static analysis alone would not be possible to extract the APIs, I have considered both static and dynamic analysis approach. Let's discuss how each of these approaches is implemented. The input for the pipeline as well as this step is an APK file of android applications. In the static analysis, I have used two tools namely apktool\cite{10} and androguard\cite{11}. In the dynamic analysis, I have mainly used the Frida instrumentation plugin.

\subsection{Static Analysis}
For a given application to be tested, the source code is not available apart from its APK file. As input, an APK file is provided. An APK (Android Package Kit) is the file format for applications used on the Android operating system. APK files are compiled with Android Studio, which is the official integrated development environment (IDE) for building Android software. An APK file includes all of the software program's code and assets. As the APK file is in compressed ZIP format, the ZIP decompression tool can open it, the best one in the market is the apktool.

\subsubsection{apktool}
apktool is used for reverse engineering 3rd party, closed, binary Android apps. It can decode resources to their nearly original form and rebuild them after making some modifications. It also makes working with an app easier because of the project like file structure and automation of some repetitive tasks like building apk, etc. It is not intended for piracy and other non-legal uses. It could be used for localizing, adding some features or support for custom platforms, analyzing applications, and much more. Some of its features include Disassembling resources to their nearly original form (including resources.arsc, classes.dex, and XMLs), rebuilding decoded resources back to binary APK/JAR, organizing and handling APKs that depend on framework resources, smali debugging (removed in 2.1.0 in favor of IdeaSmali), helping with repetitive tasks. Details on how to install the application are mentioned here: \href{https://ibotpeaches.github.io/Apktool/install/}{https://ibotpeaches.github.io/Apktool/install/}

apktool is used here in the \textit{AndroScanner} to decompress the APK file and read the permissions in the Manifest file. Android manifest file helps to declare the permissions that an app must have to access data from other apps. It also specifies the app's package name which helps the Android SDK while building the app. Every app project must have an AndroidManifest.xml file (with precisely that name) at the root of the project source set. To detail, the components added to the manifest file,

\begin{itemize}
    \item The components of the app, includes all activities, services, broadcast receivers, and content providers. Each component must define basic properties such as the name of its Kotlin or Java class. It can also declare capabilities such as which device configurations it can handle, and intent filters that describe how the component can be started.
    \item The permissions that the app needs in order to access protected parts of the system or other apps. It also declares any permissions that other apps must have if they want to access content from this app.
    \item The hardware and software features the app requires, which affects which devices can install the app from Google Play.
\end{itemize}

Using the manifest file, the permissions to the file are read to get a better idea of  app functions to be focused on. Our focus is primarily on the API calls, in order to perform API calls at the server level, an API key is used. 

\subsubsection{APIKey Extractor}
An API key is a code used to identify and authenticate an application or user. API keys are available through platforms, such as a white-labeled internal marketplace. They also act as unique identifiers and provide a secret token for authentication purposes.

APIs are interfaces that help build software and define how pieces of software interact with each other. They control requests made between programs, how those requests are made, and the data formats used. They are commonly used on Internet-of-Things (IoT) applications and websites to gather and process data or enable users to input information. For example, users can get a Google API key or YouTube API keys, which are accessible through an API key generator.

An API key is passed by an application, which then calls the API to identify the user, developer, or program attempting to access a website. It can help break development silos and will typically be accompanied by a set of access rights that belong to the API the key is associated with.

An existing tool named \textit{APIKey Extractor}\cite{12} is used in the \textit{AndroScanner} workflow to extract the API key. It searches for API keys embedded in Android String Resources, Manifest metadata, Java code (including Gradle's BuildConfig), and Native code. Further installation and usage details are provided at \href{https://github.com/alessandrodd/apk_api_key_extractor}{github.com/alessandrodd/apk\_api\_key\_extractor}.

The permissions extracted from the manifest file using the \textit{apktool} and the API key extracted using \textit{APIKey Extractor} are sent to the next step for dynamic instrumentation. In the dynamic analysis, these functions are prioritized and studied initially. Once the analysis of these functions is done, the Frida\cite{13} instrumentation plugin will be used.

\subsubsection{Androguard}

Another tool I have used in the static analysis phase is Androguard. It is primarily a Python tool to experiment with android application APK files, disassembling and decompiling Dex files (.dex - Dalvik virtual machine code),  Android's binary xml (.xml) Android Resources (.arsc). And the tool is available for Linux/OSX/Windows (python powered).

Using androguard, you can build a control flow graph and call graph using the CLI support. The call graph is constructed from the Analysis object and then converted into a networkx MultiDiGraph. Methods are nodes in the produced network, while calls are edges. The offset inside the method is kept as a property on each edge, and repeated calls between two methods result in several edges. 

The generated graphs will be enormous and give the details of all the application calls, both internal and external. Each node has an attribute to indicate if it is an internal (defined somewhere in the DEXs) or external (might be an API, but definitely not defined in the DEXs) method. The calls can be filtered further using customized scripts.

As the generated graphs are huge and require further filtering, they only provide the details of the calls but not the parameters. The parameters mentioned in the call can be extracted using dynamic analysis only.

\begin{figure*}[h]
\centering
  \includegraphics[width=0.7\textwidth,keepaspectratio]{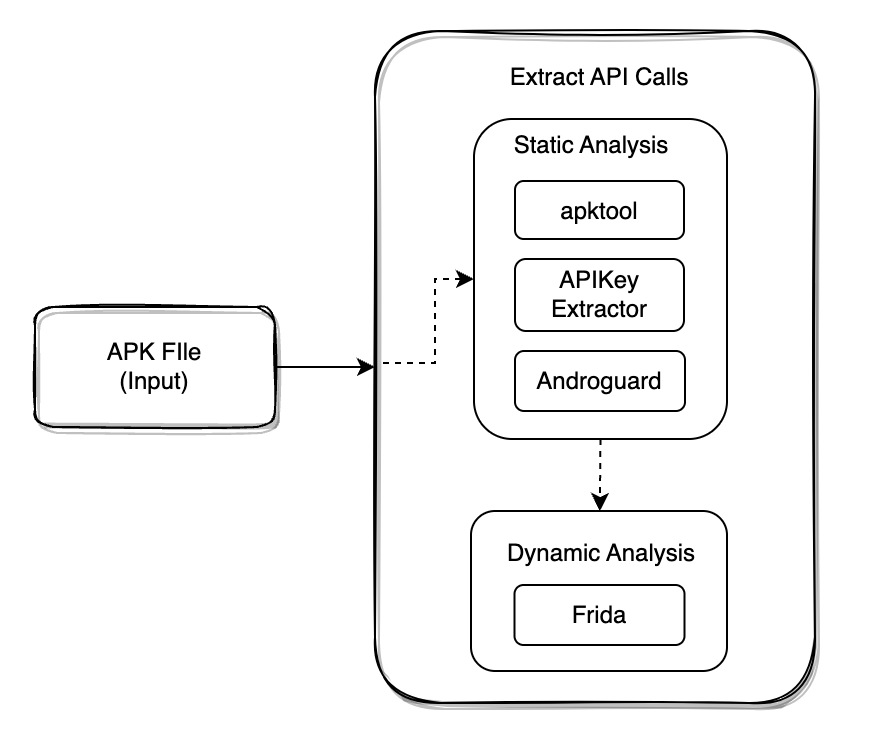}
    \caption{Extracting API Calls in AndroScanner}
  \label{ExtractAPIs}
\end{figure*}

For reference Figure \ref{ExtractAPIs} depicts the whole process involved in extracting the APIs and their parameters.

\subsection{Dynamic Analysis}
In the dynamic analysis phase, \textit{AndroScanner} uses a dynamic instrumentation toolkit \textit{Frida}. It can be used by developers, reverse engineers, and security researchers. It can incorporate custom scripts into black box process\cite{17}. No source code is required to hook any function, spy on encryption APIs, or track secret application code. The scripts can be edited, saved, and looked into the results right away. All of them can be done, without the need for compilation or application restarts.

\textit{Frida} is compatible with Windows, macOS, GNU/Linux, iOS, watchOS, tvOS, Android, FreeBSD, and QNX. Moreover, Frida is open-source software. Python and Javascript are required to access it. Python is used to communicate with the application device, and Javascript is used to hook into the desired functions.

In order to experiment with Frida, an android emulator is setup and Frida server script is run inside the android emulator (client) and javascript script to hook into the interested function calls is run on the server, in our case it is Ubuntu Linux OS.

Adb tools\cite{14} are used to connect to the android emulator/physical device. It is a powerful command-line tool for communicating with devices. The adb command simplifies device tasks such as app installation and debugging.

\begin{figure*}[h]
\centering
  \includegraphics[width=0.8\textwidth,keepaspectratio]{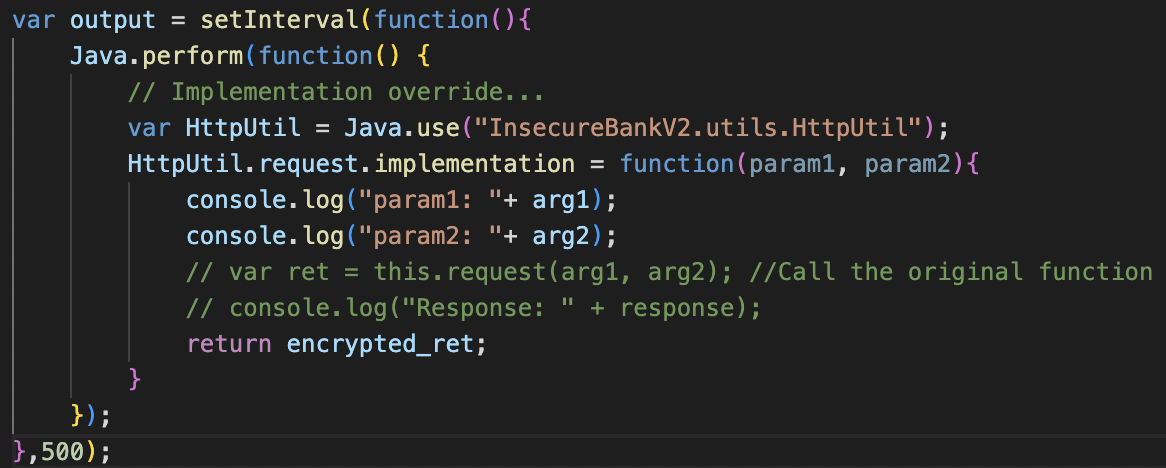}
  \caption{An example of frida hook in Javascript}
  \label{fridahook}
\end{figure*}

Frida-trace is used to hook into the functions of interest. Functions of interest are a curated list of API calls from the Android Framework API Calls which act as Network Connection calls. These calls act as entry points for the Frida-trace. Over 200 calls are collected. For example HttpParam, HttpConnectionParams, HttpCookie, etc. Some of the extracted API calls are mentioned in Table \ref{entrypoints}
While running Frida trace, using Javascript, Frida can be hooked into these entry points. The parameters used in the API calls are also extracted. An example of a frida hook using Javascript can be seen in Fig \ref{fridahook}

\begin{table*}[t]
\centering
\resizebox{0.7\textwidth}{!}{%
\begin{tabular}{lll}
\textbf{Android Framework API Entry Points} &  &                   \\
\hline                                                    \\
HttpParams                         &  & HttpCookie        \\
HttpConnectionParams               &  & HttpURLConnection \\
HttpAuthHandler                    &  & CacheRequest      \\
ContentHandlerFactory              &  & CacheResponse     \\
CookiePolicy                       &  & ContentHandler    \\
CookieStore                        &  & URL               \\
DatagramSocketImplFactory          &  & URLClassLoader    \\
HttpRetryException                 &  & URLConnection     \\
Network                            &  & URLDecoder        \\
NetworkCapabilities                &  & URLEncoder        \\
NetworkInfo                        &  & URLStreamHandler  \\
NetworkRequest                     &  & Authenticator    \\
\hline
\end{tabular}
}
% \vspace
\caption{List of Android Framework API Entry Points}
\label{entrypoints}
\end{table*}

Once the API Calls are extracted, they are categorized into internal and external calls. External calls are defined as the API services obtained from the 3rd party services like Amazon, Firebase, Apache, etc. Internal calls are defined as all the calls apart from external calls. External APIs are a curated list obtained from scanning the top 5,000 android applications from Google Playstore using LibScout\cite{15}. Over 50 of them are collected\cite{16}, they are specified in the table \ref{externalAPIs}. Using the curated list as a reference, if the extracted API contains the external library as a source that means if the API call domains contain the external library name, then it is considered as the external API Call.

\begin{table*}[t]
\centering
\resizebox{\textwidth}{!}{%
\begin{tabular}{llllllll}
{\textbf{External APIs}} &  &                  &  &               &  &              &  \\
\hline
                             &  &                  &  &               &  &              &  \\
heyZap                       &  & Fyber            &  & Stetho        &  & Bolts        &  \\
ironSource                   &  & Google           &  & Supersonic    &  & Brightroll   &  \\
jsoup                        &  & Gson             &  & Syrup         &  & Butter-Knife &  \\
roboguice                    &  & Guava            &  & Tapjoy        &  & Chartboost   &  \\
scribe                       &  & Guice            &  & Tremor Video  &  & CleverTap    &  \\
smaato                       &  & HockeyApp        &  & Twitter4J     &  & Crashlytics  &  \\
ACRA                         &  & InMobi           &  & Urban-Airship &  & Crittercism  &  \\
AMoAd                        &  & JSch             &  & Vungle        &  & Dagger       &  \\
Amazon                       &  & Joda-Time        &  & WeChat        &  & EventBus     &  \\
Segment                      &  & Pollfish         &  & AppFlood      &  & ExoPlayer    &  \\
Apache                       &  & Millennial Media &  & heyZap        &  & Facebook     &  \\
WeChat                       &  & Mixpanel         &  & ironSource    &  & Firebase     &  \\
flickrj                      &  & MoPub            &  & jsoup         &  & Flurry       &  \\
vkontakte                    &  & New-Relic        &  & roboguice     &  & AdColony     &  \\
AppBrain                     &  & OkHttp           &  & scribe        &  & AdFalcon     &  \\
AppFlood                     &  & Parse            &  & smaato        &  & Adrally      &  \\
AppsFlyer                    &  & Paypal           &  & vkontakte     &  & Fresco       &  \\
BeaconsInSpace               &  & Picasso          &  &               &  &              &  \\
                             \hline

\end{tabular}
}
\caption{List of External APIs extracted from LibScout}
\label{externalAPIs}
\end{table*}

By generating these categorizations of internal and external calls, the target user can focus on fixing the vulnerabilities in the internal calls, whereas risks corresponding to the external calls are to be reported to the appropriate 3rd party owner.

\section{Vetting the APIs}
In order to vet for the vulnerabilities in the extracted APIs, the focus is on OWASP's Top 10 API Security Vulnerabilities\cite{19}. An existing tool, APIFuzzer\cite{18} is used to analyze the API calls extracted from the previous step 'Extraction of API Calls'. APIFuzzer is a pip installable package. The tool's main features include parsing API definition from local files; JSON file input format support; all HTTP methods are supported for testing; fuzzing of request body, query string, path parameter, and request header are supported. 

APIFuzzer reads the API definition from a remote URL or local JSON file as the input. Once it analyzes, it outputs a pdf report containing the details of vulnerabilities present in a URL. The pdf document is reported to the user.

\section{Reporting the Vulnerabilities}
The pdf document containing the vulnerability details given an API URL is presented to the user as an output. As an addition, after the analysis is completed, the vulnerabilities may be disclosed to the user directly by email. This way, instead of returning to examine the generated reports frequently, the user may begin the analysis process and continue with their other duties.

\section{Experiment}
In order to test the functionality of the \textit{Androscanner} pipeline, I have taken two android applications into consideration. One is a Vulnerable Bank application and the other is currently used Google Playstore application 'Hirect'.

\subsection{Bank Application}

A vulnerable bank application apk is given as input to the \textit{AndroScanner} pipeline. The app does basic functionalities such as transferring a certain amount to another account, viewing account statements, and changing passwords for the logged-in user. Let's dive into one of the APIs, \textit{change password}, and the vulnerabilities explored.

\begin{figure*}[h]
\centering
  \includegraphics[width=0.9\textwidth,keepaspectratio]{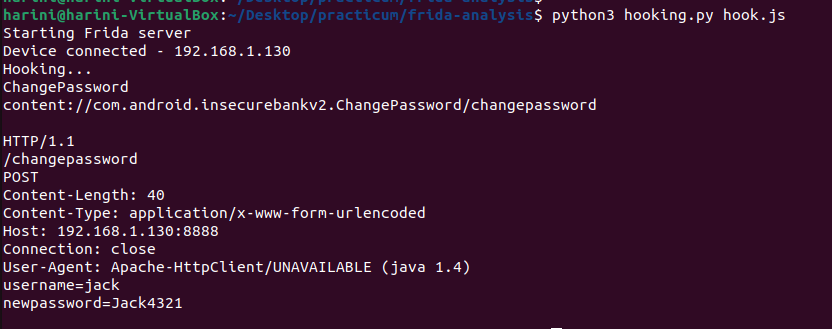}
  \caption{Extracting changepassword API Call in Bank Application}
  \label{changepassword}
\end{figure*}

Although androguard from the static analysis did not provide any useful information, after the function hook in the dynamic analysis, frida prints out the function and its parameters corresponding to the call. After formatting, the generated parameters can be seen in the Figure \ref{changepassword}.

\begin{figure*}[h]
\centering
  \includegraphics[width=0.9\textwidth,keepaspectratio]{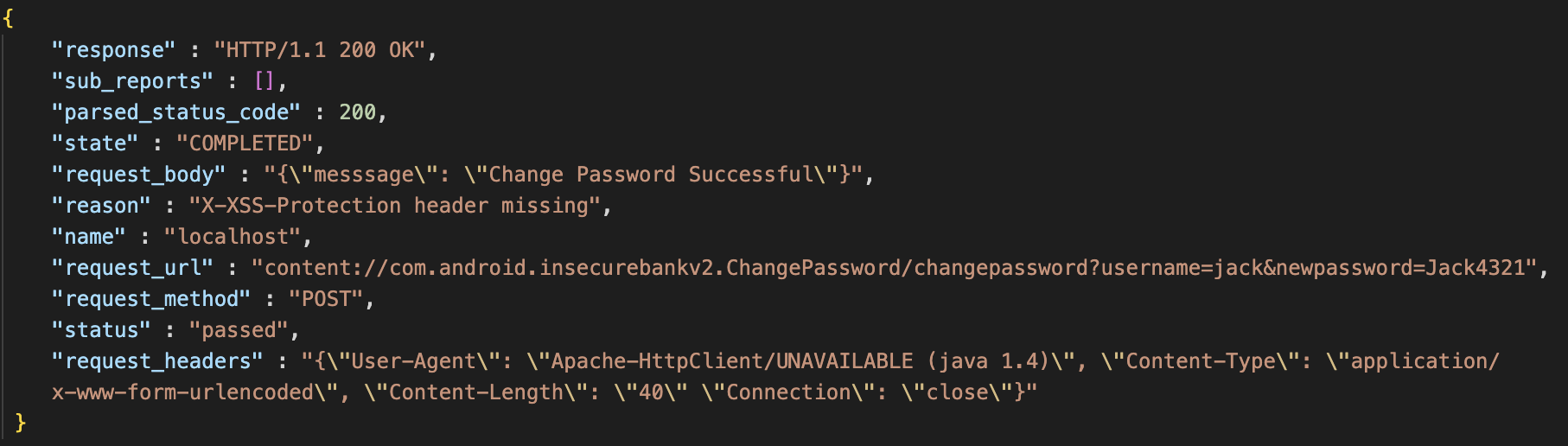}
  \caption{Reported Vulnerability by APIFuzzer for changepassword}
  \label{reportcp}
\end{figure*}

Once the API and its parameters are obtained, json format of the API is passed to the APIFuzzer. It then reports the vulnerabilities in the given API call which can be seen in Figure \ref{reportcp}

It reports that the HTTP X-XSS-Protection response header is missing which is the most basic header that prevents cross-site scripting.

\section{Results}

\begin{figure*}[h]
\centering
  \includegraphics[width=0.8\textwidth,keepaspectratio]{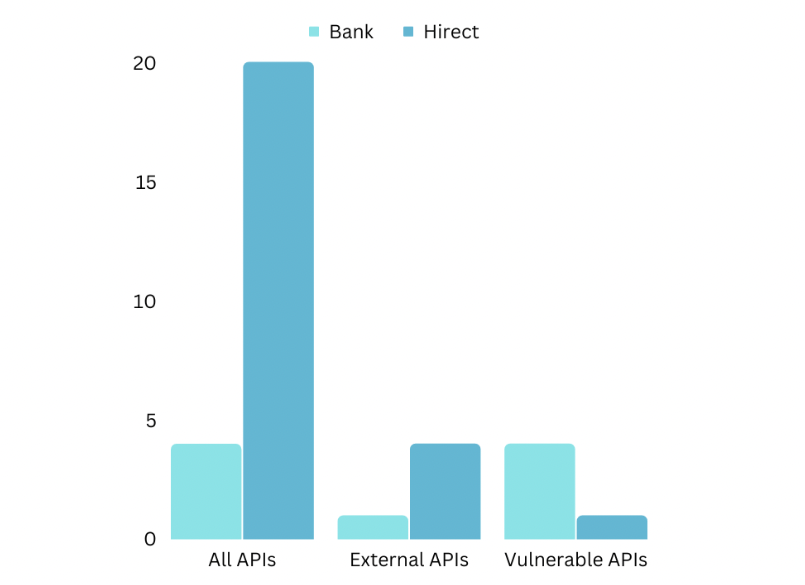}
  \caption{Comparison of Extracted APIs for Bank and Recruitment Applications}
  \label{results}
\end{figure*}

After running both the applications into the \textit{AndroScanner}, a list of vulnerabilities obtained can be compared in Figure \ref{results}

All the APIs that could be extracted from the Bank and Recruitment application are 4 and 20 respectively. Out of which, the external APIs obtained using Table \ref{externalAPIs} is 1 and 4 respectively. The other 3 and 16 APIs are internal APIs. The vulnerabilities exposed in the Bank and Recruitment applications are 4 and 1 respectively.
\hfill \break
The domains for the extracted APIs for the Bank and Recruitment app are mentioned in the Table \ref{resultAPIs} 

\begin{table*}[t]
\centering
\resizebox{0.85\textwidth}{!}{%
\begin{tabular}{|c|l|l|}
\hline
\textbf{API}                & \multicolumn{1}{c|}{\textbf{Bank App}} & \multicolumn{1}{c|}{\textbf{Recruitment App}} \\ \hline
\multirow{4}{*}{external}   & \multirow{4}{*}{fonts.gstatic.com}     & settings.crashlytics.com                 \\ \cline{3-3} 
                            &                                        & e.crashlytics.com                        \\ \cline{3-3} 
                            &                                        & api.wechat.com                           \\ \cline{3-3} 
                            &                                        & bcdn.wechat.com                          \\ \hline
\multirow{4}{*}{vulnerable} & insecurebankv2.ChangePassword          & \multirow{4}{*}{seekermsg.hirectapp.com} \\ \cline{2-2}
                            & insecurebankv2.PostLogin               &                                          \\ \cline{2-2}
                            & insecurebankv2.DoTransfer              &                                          \\ \cline{2-2}
                            & insecurebankv2.ViewStatement           &                                          \\ \hline
\end{tabular}
}
\caption{External and Vulnerable APIs recorded in the Bank and Recruitment App}
\label{resultAPIs}
\end{table*}
Hirect is an android application that focuses on employment-oriented online service that operates via websites and mobile apps. The vulnerability found in the seekermsg endpoint is Excessive Data Exposure, which is ranked 3rd among the OWASP API Security Vulnerabilities. It discloses the timestamps of messages sent to users. An attacker can access this timestamp to generate a custom message pretending to be a recruiter and exploit the job seeker. This vulnerability was responsibly disclosed to the Hirect development team prior to publication of this paper. Further technical details are withheld pending remediation.

Although I have found one vulnerability in the recruitment application, I expected it to be zero as it is one of the popular apps used in today's applications. This motivates me to scan the other top android applications from PlayStore as part of my future work.

\textit{Challenges Covered:}
One of the challenges faced in any of the tools nowadays is customized functions that are built by the application developers to perform the network connection calls. The present pipeline only detects the functions included in the curated Android Framework entry points list. This can be solved by an extension that takes in customized javascript functions as Frida hooks to analyze. All the scripts under a 'customize' folder are provided as function hooks during dynamic analyses.

I have also contacted an Android Developer - Sonu Saurab who works at ShareChat to review the \textit{AndroScanner}. He has mentioned that the application could extract 80 percent of the API calls as far as the scope taken. However, he could not understand the reported vulnerabilities from the pipeline and the actions to be performed. He mentioned that GUI support for the tool could help him better comprehend the vulnerabilities and patch them. He also sought an IOS-compatible tool.

\section{Target Users}
The target users for the project are the app developers working on the application backends, researchers, or analysts working in the android application Pen-testing area. The pipeline serves as black-box testing in the pre-production phase.

The pipeline \textit{Androscanner} can also be used as Platform-as-a-Service (PaaS) to test the mobile applications at scale. However, the current pipeline requires some tweaks such as increased CPU performance, to undertake testing at large volume.

\section{Limitations \& Deployment}
\subsection{Limitations}
\subsubsection{Android only compatible}
The current pipeline \textit{Androscanner} is compatible only with android APK files and these files must be JAVA based. Hence, It does not support IOS applications.

\subsubsection{Encrypted Parameters}
Frida hooks into the function calls and extracts the API call parameters. If these parameters are encrypted using the customized encryption function, there is no way for the \textit{AndroScanner} pipeline to detect these parameters automatically. As the encryption function is not included in the entry points functions to hook, the scanner would not detect it and extracted parameters would be still encrypted.

\subsubsection{CLI Support}
As discussed in the implementation section, the pipeline is accessible using the Linux Command Line Interface support. The process of interacting with tools is automated using bash scripts.

\subsection{Deployment}
\textit{Androscanner} pipeline is hosted on Linux Operating System (Ubuntu 22.04). Major softwares installed are as follows:
\begin{itemize}
    \item Python
    \item Java
    \item apktool
    \item APIKey Extractor
    \item Androguard
    \item Android Emulator
    \item Adb tools
    \item Frida
    \item Javascript
    \item APIFuzzer
\end{itemize}

\section{Conclusion}
\textit{AndroScanner}, an analytical pipeline for studying mobile app backends, was introduced in this paper. I have analysed empirically two of the applications, one is a Vulnerable Bank Application and the other is a production recruitment application with over 50k+ downloads on Google Playstore. The vulnerable Bank Application is used to test the pipeline and fine-tune it. It has discovered a zero-day vulnerability in the recruitment application. Finally, \textit{AndroScanner} is available upon request to assist app developers in improving the security of their backends, providing insight into which platforms are susceptible, and guiding developers in resolving vulnerabilities discovered in their backends.

\section{Future Scope}
The following extensions can be considered as part of the future work to develop the pipeline.

\subsection{Confidence score}
Currently, \textit{Androscanner} only uses APIFuzzer to extract vulnerabilities, given an API URL. Instead of relying on one tool, multiple tools can be used to extract vulnerabilities. For a given URL, the extracted vulnerabilities corresponding to single technology can be bundled up and a confidence score can be computed. The confidence score of a vulnerability indicates the confidence or likelihood that the vulnerability is affecting the target URL for an application. Thus, the confidence score gives an extra layer of understanding the vulnerabilities present in the application.

\subsection{Vulnerability Ranking}
In the given project, the target users are mainly app developers. The output of the third step of the \textit{Androscanner} lists the vulnerabilities existing in a URL. The problem here is, as the app developer is not security-focused, they would not understand which of these listed vulnerabilities are important. They cannot prioritize the risks. Hence, using an existing vulnerability database such as \href{https://www.exploit-db.com/}{ExploitDB} can be used to rank the vulnerabilities. They can be categorized as 'High', 'Medium', and 'Low' severity. This gives the developer a clear picture of risks.

\subsection{Patch Suggester}
Vulnerability ranking functionality can be extended to provide a patch suggester to fix the vulnerabilities listed. Existing patch suggesters developed by Syxsense\cite{20} which is the best in the market can be used. However, it is not an open-source tool. Anyway, building a patch suggester from scratch has its own challenges. One of the solutions would be developing using machine learning models.

\subsection{Automatic Reporting}
Considering the current \textit{Androscanner} pipeline, the list of vulnerabilities is reported as a pdf to the target user. If multiple applications are being tested at a go corresponding to different owners, the reporting is difficult. Hence, a backend functionality that stores the owners' details like email, and phone number in the database (MySQL) and reports the vulnerabilities automatically to the owner is needed. It can be built using Python and MySQL.

\subsection{Docker Container}
The present pipeline is deployed on a Ubuntu Linux Operating System with a number of required software packages installed for running the tools. An app developer may have to use distinct versions of the same software for various projects. To solve this problem, docker containers can be used. Docker streamlines the development lifecycle by allowing developers to work in standardized environments using local containers which provide your applications and services. Containers are great for continuous integration and continuous delivery (CI/CD) workflows.

\subsection{GUI Support}
As discussed previously regarding the feedback from the Android Developer, one of the requirements mentioned is GUI support. This would allow the app developer to easily navigate through the tools and use them effectively without worrying about the contents of the tools.

\section*{Acknowledgment}
The author would like to thank Prof. Mustaque Ahamad of the Georgia Institute of Technology for his guidance and support throughout this work. This work was originally completed in December 2022 and is being made publicly available here.

\end{document}